\documentclass[sn-mathphys]{sn-jnl}

\usepackage{graphicx}%
\usepackage{multirow}%
\usepackage{amsmath,amssymb,amsfonts}%
\usepackage{amsthm}%
\usepackage{mathrsfs}%
\usepackage[title]{appendix}%
\usepackage{xcolor}%
\usepackage{textcomp}%
\usepackage{manyfoot}%
\usepackage{booktabs}%
\usepackage{algorithm}%
\usepackage{algorithmicx}%
\usepackage{algpseudocode}%
\usepackage{listings}%
\usepackage{graphicx}
\usepackage{adjustbox}
\usepackage{array}
\usepackage{booktabs}

\theoremstyle{thmstyleone}%
\theoremstyle{thmstyletwo}%

\theoremstyle{thmstylethree}%

\begin{document}

\title[Article Title]{Visual Harmony: 

Text-Visual Interplay in Circular Infographics}


\author[1]{\fnm{Shuqi} \sur{He}}
\author[1]{\fnm{Yuqing} \sur{Chen}}
\author[1]{\fnm{Yuxin} \sur{Xia}}
\author[1]{\fnm{Yichun} \sur{Li}}
\author[1]{\fnm{Hai-Ning} \sur{Liang}}
\author*[1]{\fnm{Lingyun} \sur{Yu}}\email{lingyun.yu@xjtlu.edu.cn}

\affil*[1]{\orgname{Xi'an Jiaotong-Liverpool University}, \orgaddress{\city{Suzhou}, \state{Jiangsu}, \country{China}}}


\abstract{Infographics are visual representations designed for efficient and effective communication of data and knowledge. One crucial aspect of infographic design is the interplay between text and visual elements, particularly in circular visualizations where the textual descriptions can either be embedded within the graphics or placed adjacent to the visual representation. While several studies have examined text layout design in visualizations in general, the text-visual interplay in infographics and its subsequent perceptual effects remain underexplored. To address this, our study investigates how varying text placement and descriptiveness impact pleasantness, comprehension and overall memorability in the infographics viewing experience. 
We recruited 30 participants and presented them with a collection of 15 infographics across a diverse set of topics, including media and public events, health and nutrition, science and research, and sustainability. The text placement (embed, side-to-side) and descriptiveness (simplistic, normal, descriptive) were systematically manipulated, resulting in a total of six experimental conditions.
Our key findings indicate that text placement can significantly influence the memorability of infographics, whereas descriptiveness can significantly impact the pleasantness of the viewing experience. Embedding text placement and simplistic text can potentially contribute to more effective infographic designs. These results offer valuable insights for infographic designers, contributing to the creation of more effective and memorable visual representations.}




\maketitle

\section{Introduction}\label{sec:introduction}

Infographics, which represent data and information through a combination of text and schematics, have been increasingly utilized across various domains such as education \cite{Bicen_2017,Kibar_2014,Davidson_2014,Gebre_2016}, business \cite{Young_2014}, and healthcare communications \cite{McCrorie_2016, Scott_2016, Martin_2019}. They have been regarded as effective and visually engaging tools for communicating and simplifying complex ideas into easily digestible, bite-size representations \cite{borkin2013makes,Zhu_2020}. Their ability of presenting information in a compelling and concise way is particularly helpful in storytelling with an overwhelming amount of data in this era of information overload \cite{metoyer2018coupling, Barral_2020}. Consequently, growing research interests have emerged in the field of visualization concerning how to create an effective infographic design \cite{gay2019audit}.

    A well-designed infographic is much more than just a pretty picture. Rather, it involves thoughtful and strategic considerations of various design elements such color, icons and typography. The use of color in infographics, for example, goes beyond aesthetic appeal. Strategic color choices can effectively differentiate between information sections, enhancing the readability and understanding of the visualization \cite{borkin2013makes}. Likewise, icons and other graphical elements can capture viewer's attention and significantly improve communication efficiency when thoughtfully selected and employed \cite{Haroz_2015, Munzner_2014, metoyer2018coupling}. Finally, typography, which concerns the style, appearance, and arrangement of text, is another essential element in infographic design. A careful selection of fonts and text styles can enhance the legibility and aesthetics of visualizations. Moreover, typographic emphases, such as bolding, italicizing, and underlining, can highlight critical data points or messages in a visualization \cite{Walker_2000}. In this context, text does not only serve a structural purpose, but it can also communicate subtle nuances in the narrative that may not be expressed through the graphical elements alone.

Individual design elements such as color, icons, and typography undeniably play a significant role in infographic design, as supported by extensive research in the field. Studies have indicated the influence of well-selected color palettes in enhancing information representation and addressing cultural nuances \cite{arcia2019colors,Lucius_2017}. The impact of icons and visual embellishments on aspects like memorability, visual search, and comprehension is also well documented \cite{Borgo_2012,Skau_2015}. Furthermore, research into typography has shed light on how variable font variations can contribute to encoding data into readable text \cite{Lang_2022}.

However, while individual elements are crucial, the integration and arrangement of these elements into a holistic infographic is equally, if not more important. After all, infographics are not a random assortment of visuals and texts, but crafted through an integrated representation that results from the deliberate and harmonious interplay of these components. While receiving research interests in interactive web-based document design \cite{latif2018exploring,latif2021deeper}, the effect of text-visual interplay remains relatively under-explored within the specific context of infographics. As far as we know, there are few studies that have empirically investigated the integration of these elements within infographics. Therefore, this study aims to examine how the arrangement and interaction of text and visuals can contribute to effective infographic design. To simplify the research landscape and offer a more targeted exploration, we have chosen to narrow our scope to standalone data visualization infographics, specifically those involving \textbf{circular visualizations}. This decision was guided by two main factors. Firstly, circular visualizations are widely used and account for 69\% of glyphs in a comprehensive corpus that combines images from multiple visualization publications~\cite{Deng:2020:VisImages}. Secondly, the structure of circular visualizations provides ample opportunities for text embedding, making them an ideal candidate for studying the interplay between text and visuals. Consequently, we investigate how texts of varied levels of descriptiveness can be placed in relation to the visuals in these circular infographics. 

\begin{figure}[bp]%
\centering
\includegraphics[width=\linewidth]{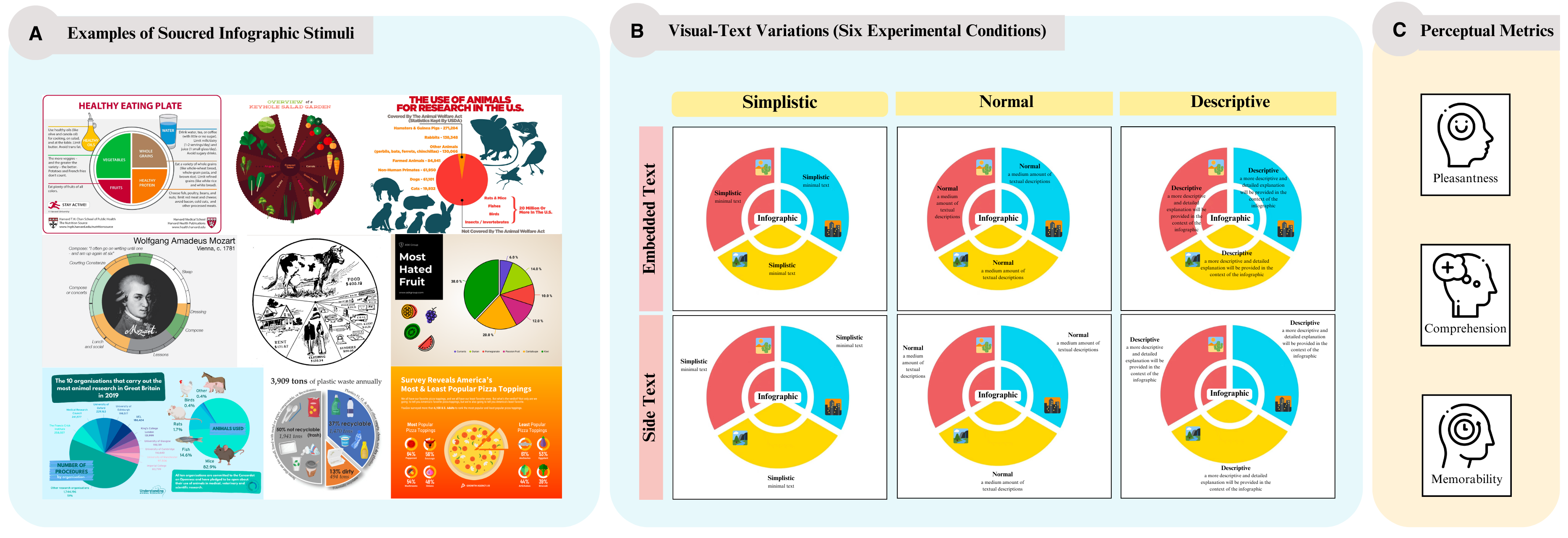}
  \caption{Overview of the study: (A) a selection of diverse infographic stimuli used as a starting point for our study; (B) Modifications to each infographic, altering text placement and descriptiveness to form six distinct experimental conditions; (C) The consequential assessments of these variations, measured through perceptual metrics such as pleasantness, comprehension, and memorability.}
	\label{fig:teaser}
\end{figure}

This focus on the text-visual interplay within circular visualizations leads us to the broader question of how design choices in infographics impact user perception. After all, the effectiveness of an infographic heavily relies on how it is perceived by viewers. In other words, user experience and perception are key indicators of infographic quality. In light of this, our study employs three dimensions of perceptual metrics to assess the effectiveness of design. These metrics, namely pleasantness, comprehension, and memorability, each measure distinct aspects of infographics performance. \textbf{Pleasantness} serves as a measure of the viewer's overall viewing experience. \textbf{Comprehension}, on the other hand, probes the success of information transfer from the infographic to the viewer, assessing how effectively the data and knowledge are being communicated. Lastly, \textbf{memorability} reflects the longevity of information retention, providing insights into whether the presented information can be remembered by the viewer. Collectively, these metrics help quantify the perceptual effects of our text-visual variations.

In this study, we collected a set of 15 circular infographic visual stimuli sourced from various websites and created six distinct variations for each infographic with two different \textit{text placements} (\textbf{embedded text, side text}) and three levels of \textit{text descriptiveness} (\textbf{simplistic, normal, descriptive}). We recruited 30 participants and assessed their visual perception with questionnaires tailored to the infographic stimuli, probing their \textit{pleasantness}, \textit{comprehension} and \textit{memorability}. \autoref{fig:teaser} serves as a step-by-step visual representation of our comprehensive investigation into the impact of text-visual interplay in infographics.

    Our study contributes to the current literature of infographics design in three aspects. Firstly, it extends our understanding of text-visual interplay in the context of infographics, an area which, despite its importance, has been relatively under-explored in the existing literature. By investigating how the descriptiveness and the placement of text can influence the viewer's experience, our study offers a perspective on the intricate dynamics between text and visuals in infographic design. Secondly, our study provides evidence-based design implications for creating visually-pleasing and memorable infographics. The perceptual metrics of pleasantness, comprehension, and memorability shed light on how strategic manipulation of text placement and descriptiveness can optimize the viewing experience and enhance information retention. Finally, our work can serve as an inspiration for future research in this field. The findings of this study encourage further exploration into the integration of various design elements in infographics. Our focus on the interaction between text and visuals is just one facet of this complex design puzzle. We hope that our efforts will spark more comprehensive explorations of the interplay among different design components and in different types of infographics.

\section{Related Work}
\label{sec:related}

\subsection{Circular Visualization}

Geometric abstraction, encompassing shapes such as lines, circles, and triangles, has a long history of being used as a visual strategy in data representation. Serving as both tools for data illustration and elements of visual interests, these shapes enable the design of informative and visually appealing infographics \cite{Horn2000, Magagnini20}. 
Among these geometric forms, the circle holds particular significance due to its expressive potential and extensive connections to everyday objects. This circle's appeal was profoundly captured by Wassily Kandinsky, one of the pioneers of abstract art. Kandinsky remarked, the circle ``is the most modest form, but asserts itself unconditionally,'' and it is ``a precise, but inexhaustible variable \cite{Islam_2021}.'' This powerful symbolism of circles, mirrored in Kandinsky's own influential abstract works, combined with its geometric simplicity, makes it an attractive and meaningful element in infographic design. This potential was recognized early on by pioneers such as Playfair \cite{Spence05}, with the creation of the pie chart, and Nightingale \cite{Nightingale_1999}, who developed eccentric circular visualizations for analyzing the causes of death of the British army during the Crimean war.

Over time, circular visualizations have evolved and diversified to encompass various visual models such as rings, spirals, wheels, and even flow diagrams involving grids and maps \cite{Lima2017}. Their adaptability in representing complex data relationships and interconnected dynamics has made them a preferred model in infographic design. For instance, in Magagnini’s work \cite{Magagnini20}, circular visualizations are chosen to visualize and organize contemporary art data. They can also be used to represent categories, proportions, or cyclic time-dependent data \cite{draper2009survey, cui2021mixed}. For example, Buono et al. \cite{buono2014circular} utilized a circular stripe design to visually represent collaborative team activities.

Among circular visualizations, a subgroup employs a space-filling design pattern, where the areas within the circular layout are densely filled \cite{draper2009survey}. A unique feature of these space-filling circular visualizations is their capacity to incorporate and encode textual information, thereby providing designers with the unique opportunity to creatively blend text and visual elements. This fusion of text and visuals lays the groundwork for the exploration of how text placement and descriptiveness might impact the effectiveness of infographics, a question we delve into in the subsequent sections of this paper.

\subsection{Text-Visual Interplay}
The interaction between text and visuals has been studied across different domains from areas like web-based storytelling, document reading, and narrative visualizations.

The impact of text-visual layout and linking has been investigated in web-based storytelling. For instance, Zhi et al. \cite{Zhi_2019} revealed that in stories containing both text and visualizations, the choice of layout, specifically the comparison between vertical and slideshow formats, significantly impacted the reader's comprehension and overall experience. Interactive linking, the explicit connection between narrative text and its corresponding visualization, was found to improve user engagement, enhance comprehension, and mitigate layout-related challenges.

Another area of focus has been the integration of text and visuals in adaptive documents. For instance, Badam et al. \cite{Badam_2019} worked on enhancing comprehension of data-rich documents by dynamically generating on-demand visualizations based on the reader's current focus. Their approach highlighted the potential of tight coupling between text and visuals in enhancing understanding. On a related note, Barral et al. \cite{Barral_2020} designed an adaptive guidance system for processing narrative visualizations embedded within textual documents. They demonstrated that guiding users' attention to salient components in the visualization, especially through transitions, could significantly improve comprehension.

In the context of infographics, previous studies have predominantly focused on the role of text and visuals in visualizations involving information flows and timelines. For instance, Lu et al. \cite{Lu_2020} developed the concept of Visual Information Flow (VIF), where visual elements are linked together to convey information and narrative. However, their research did not delve into the specific interplay between textual and visual components. Majooni et al. \cite{majooni2018eye} compared the effectiveness of different arrangements such as the zigzag form and other alternatives in infographic structures, providing evidence that layout significantly influences viewer comprehension and cognitive load. Despite these advancements, the specific text-visual interplay and placement in flat infographics, especially those featuring circular visualizations, remain underexplored.

\subsection{The Role of Perceptual Metrics}
The role of text and visuals extend beyond mere strategies of integration; it also significantly impacts the viewer's perception of the presented information. Consequently, perceptual metrics serve as important performance indicators for different infographic design.

In this study, we focus on three key dimensions of perceptual metrics: Pleasantness, a measure of the viewer's overall viewing experience. Comprehension, a metric that probes the efficiency of information transfer from the infographic to the viewer. Memorability, a measure of the viewer's ability to retain the presented information over time. These three dimensions together holistically quantify the perceptual effects of varying text-visual interplays in infographics, offering valuable insights into the creation of effective designs.

\textbf{Pleasantness} is an example of affective response that can intertwine with the engagement experience of data visualizations.  In many instances, feelings such as pleasantness can mirror the depth of user engagement and the overall quality of the viewing experience \cite{Kennedy_2018}. The interplay between affective responses and infographics has been examined in previous research. For instance, Lan et al.'s research \cite{Lan_2021} investigated this intersection and revealed that viewers can form consistent judgments on the affective responses elicited by infographics. However, our study's perspective on pleasantness is not geared towards the exploration of emotions provoked by the infographics. Instead, we concentrate on the direct viewing experience of the infographics themselves. This approach aligns with the efforts of other researchers, including McGurgan et al. \cite{mcgurgan2021graph}, who assessed aesthetic pleasantness through the lens of expert users, examining charts with varying levels of data ink. 

\textbf{Comprehension} plays a central role in infographic design as it pertains to the viewer's mental processing of information. It therefore serves as the cognitive bridge between the data presented in the infographic and the viewer's knowledge, facilitating an enhanced understanding of the material \cite{Albers_2015}.
Various studies have highlighted the use of comprehension as a perceptual metric in infographic related research. Gallagher et al. \cite{Gallagher_2017} used comprehension as a perceptual metric for investigating the usefulness of infographics in summarizing information in online learning. They measured comprehension by asking learners to rate on a five-point Likert scale whether the infographics were helpful in summarizing and elucidating learning materials. Meanwhile, Young et al. \cite{Young_2014} compared infographics with their text equivalents. Their evaluation of comprehension involved the use of multiple-choice questions related to the content of the article, with correct answers earning points. Similarly, Alsaadoun’s \cite{Alsaadoun_2021} research delved into the differential comprehension between infographic-aided instructions and traditional text instructions. They also gauged comprehension using a multiple-choice test, the results of which suggested that infographics can notably enhance comprehension. Furthermore, Obie et al. \cite{obie2020effect} found that the use of author-driven narratives in visualization could boost both memorability and comprehension. Their study used Hans Rosling's charts and found that participants who interacted with the author-driven videos had better recall and comprehension than those who interacted with the standalone interactive visualizations.

\textbf{Memorability} is another significant factor in evaluating the effectiveness of data visualizations. The ability of a viewer to recall the information presented in an infographic or visualization is a key determinant of its success.
The study conducted by Borkin et al. \cite{borkin2013makes} highlighted the use of memorability as a broad metric for the utility of information, underlining that viewers consistently identified certain visualizations as memorable. In their findings, unique visualization types outperformed common graphs in memorability. 

Meanwhile, the exploration of data storytelling and its effects on memorability has led to intriguing findings. Wang et al. \cite{Wang_2019} and Zdanovic et al. \cite{Zdanovic_2022} conducted studies focusing on the role of narrative and storytelling in data visualization. They found that data comics and storytelling visualizations, respectively, improved understanding and the ability to recall information. Contrastingly, Gareau et al.'s research \cite{gareau2015exploration} did not find a significant memory-related advantage for infographics over traditional text documents when visualizing Census data. However, they noted an interaction between the type of stimulus and the nature of the task (searching versus recall), with infographics improving performance in the searching task rather than recall tasks.
Drawing upon these various studies, our investigation aims to assess memorability as a perceptual metric in our evaluation of text-visual interplay in infographics.

\section{Method}

\begin{figure*}[ht]
  \centering
  	\includegraphics[width=\textwidth]{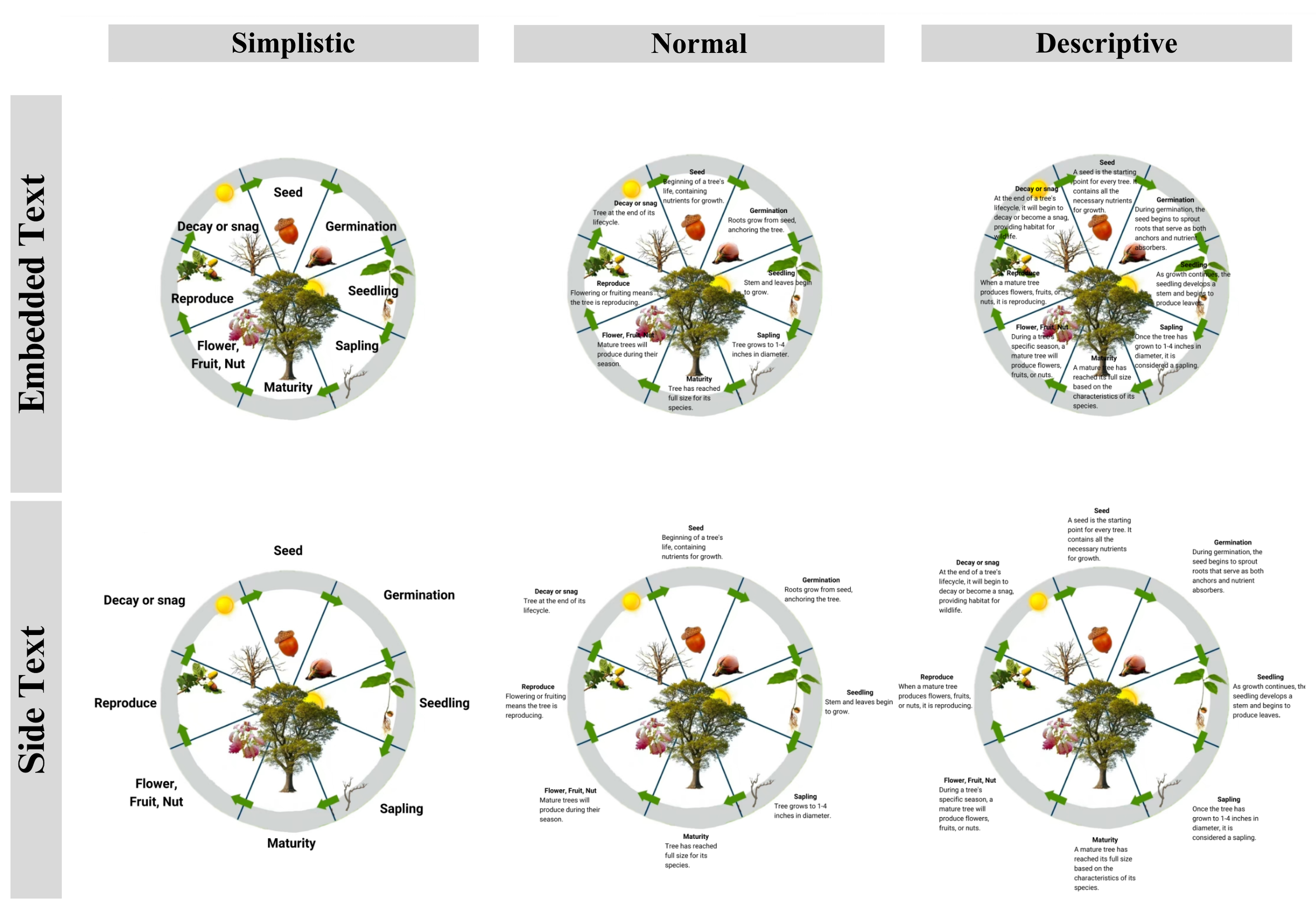}
  	\caption{An example infographic stimuli edited to six versions with varied text placement and descriptiveness levels. This particular infographic captures information about the life cycle of trees \cite{farmcredit2022treelifecycle}. Six distinct versions of the same infographic have been created. Each version systematically varies in terms of where the text is placed (embedded text vs side text) and the level of descriptiveness provided (ranging from minimalistic to descriptive).}
  	\label{fig:example}
   \vspace{-2mm}
\end{figure*}

\subsection{Participants}
We recruited a total of 30 participants (16 female, 14 male) through direct contact. The age of the participants ranged from 21 to 27, with an average age of 24 for the study. The participant pool consisted of 5 undergraduate students, 23 master's students, and 2 PhD students and have diverse backgrounds in fields such as business and finance (8), linguistics (1), education (1), mathematics (1), design (1), and computer science (18). The participants' level of experience interacting with circular visualizations was screened and recorded. It is noteworthy that 90\% of the participants had previous experience with circular visualizations. Among these individuals, 3 interacted with circular visualizations annually, 10 on a weekly basis, 4 on a daily basis, and 2 multiple times a day. This sample size was selected to ensure a diverse range of perspectives while keeping the data manageable. The participants were grouped into 6 groups, corresponding to the 6 experimental conditions. 
Before initiating the experiments, participants were provided with background information about the project and voluntarily gave their consent to participate in the data collection process.

\subsection{Infographic Stimuli Selection}
To provide the participants with realistic examples of infographics, we carefully curated a set of 15 infographics from various websites \cite{balliett2011dos, brinton2016ye, healthyeatingplate, faunalyticsinfographics, abpbehavioralmentalhealth, pinterest2019realpiechart, uar2020animalresearch,pinterest2022keyholegarden, venngageexplodedpiechart, farmcredit2022treelifecycle,infowetrustroutines,engagecupertino2021plasticsdata, venngagesurveyresultsinfographic, venngagewastemanagementinfographic, venngagedarkghgchart}. Three factors were considered when selecting appropriate infographics. First, the infographic must feature a circular visualization. Second, the infographic should present an acceptable level of clarity and coherence in its data and information. Third, we sought to exclude infographics containing obscure, domain-specific concepts that might be beyond the understanding of our target participant pool. Apart from these, we avoided overly simplified infographics, which offered no scope for text modification into more descriptive statements.

The curated collection of infographics all featured circular data visualizations and covered broad spectrums of themes, including media and public events, health and nutrition, science and research, and sustainability. This selection was intentionally diverse to represent different domains of information, hence offering a well-rounded sample of infographic stimuli for our study.

\subsection{Experimental Conditions}

In this section, we detail the process of modifying each of the selected infographic stimulus to produce varied versions that served as our study's experimental conditions. 

These strategic alterations allowed us to systematically examine the effects of text-visual interplay.

First and foremost, we define the distinctions between the six experimental conditions in terms of text placement and text descriptiveness. For text placement, we explored two commonly used methods in circular visualizations: embedding the text within the graphics and placing the text side-by-side with the visual representation. For descriptiveness, we categorized the text into three types: simplistic, normal, and descriptive. These categories were defined based on both the length and complexity in the text: the simplistic condition contains fewer than 5 words, the normal condition contains around 10 words, and the descriptive condition contains around 20 words. These conditions provided a comprehensive overview of the possible combinations of text placement and descriptiveness in circular visualizations.

Then, we modified each selected infographic to create a set of six varied versions, corresponding to our experimental conditions. First, we extracted the text from the original infographics while keeping the visual elements intact. We then altered the descriptiveness of the initial textual information through processes such as rewording, rephrasing, shortening, expanding, and reorganizing. This process yielded three distinct versions of text, varying in length but maintaining the same core message. The quality of the texts, the consistency with our pre-defined descriptiveness levels, and the preservation of the core information were validated by an external evaluator. Lastly, we reassembled the varied infographics by reintegrating the corresponding text versions into the visual elements. Depending on the version, we either embedded the text within the graphic or placed it alongside the visual representation. \autoref{fig:example} illustrates an example set of the six variations produced from a single original infographic.

\subsection{Presentation of Infographic Stimuli}

The infographic stimuli were presented in a digital format on the screen of a 13-inch personal computer. 

Since text descriptiveness is a within-subject factor, each participant was presented with an equal proportion of simplistic (5), normal (5) and descriptive (5) infographics. To control for order effects and counteract the differences in distinct infographics, we adopted a Latin square design, ensuring that the conditions were counterbalanced.

\subsection{Experimental Procedure and Evaluation Methods}

In this section, we provide details on the empirical experiments conducted with the participants and the collection of accuracy and perceptual metrics. 

\begin{figure}[htb]
  \centering
  \includegraphics[width=0.5\columnwidth]{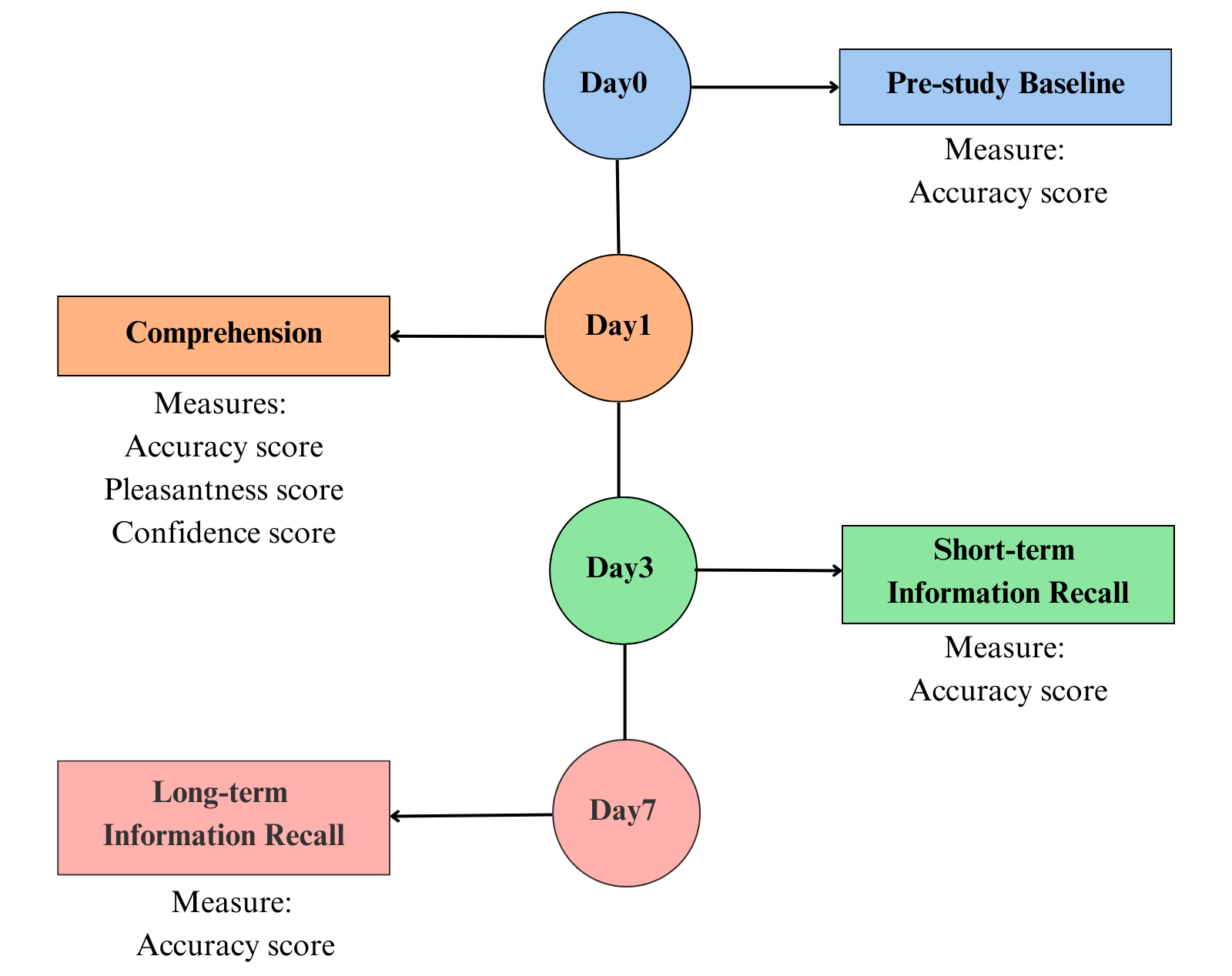}
  	\caption{Illustration of the experimental workflow.}
  	\label{fig:workflow}
\end{figure}

Our experiment consisted of four distinct sessions: pre-study baseline, comprehension, short-term information recall, and long-term information recall. 
This approach draws inspiration from Zdanovic et al.'s \cite{Zdanovic_2022} method of assessing memorability at three different stages, which we have enhanced by incorporating an additional testing phase three days after the presentation of stimuli to differentiate between short- and long-term memorability. Similar multi-phase evaluation methods have been employed in previous studies, such as Obie et al.'s \cite{obie2020effect}, which used multiple choice questions to evaluate information recall across different intervals. The duration of our experiment sessions varied: the pre-study, short-term, and long-term recall sessions each lasted approximately 10 minutes, while the comprehension assessment ranged from 20 to 50 minutes, depending on the participant's pace. The overall workflow of the experiment sessions is illustrated in \autoref{fig:workflow}.

\begin{figure}
  \centering
  \includegraphics[width=0.5\columnwidth]{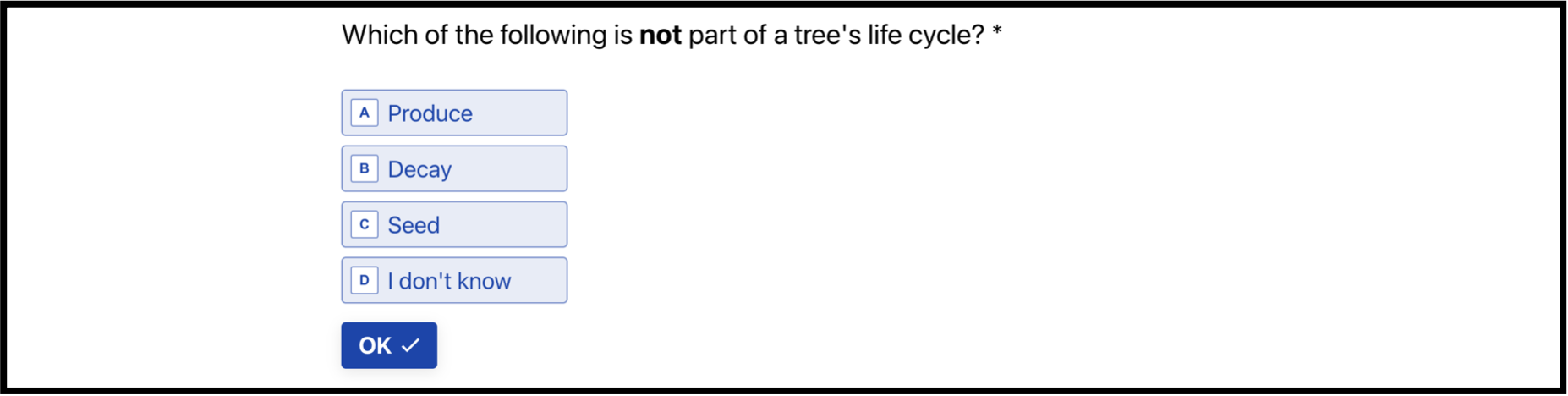}
  \caption{Pre-study baseline assessment Interface.}
  \label{fig:baseline}
\end{figure}

  \begin{figure}
  \centering
  \includegraphics[width=0.5\columnwidth]{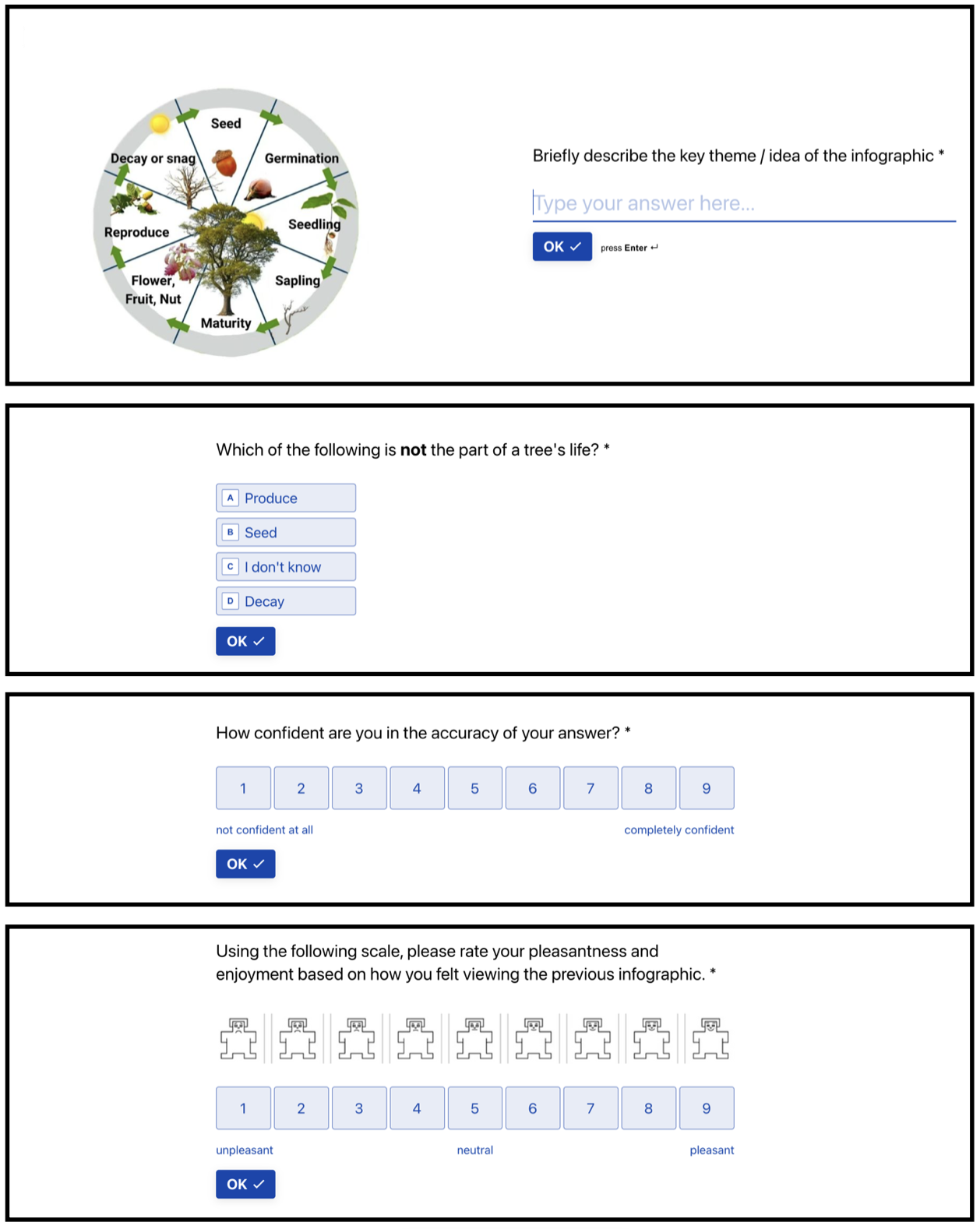}
  \caption{Comprehension assessment interface showcasing four sequential frames: (1) infographic viewing and summarizing, (2) answering a multiple choice question, (3) reporting the confidence score, (4) reporting the pleasantness score.}
  \label{fig:comprehend}
  \end{figure}

  \begin{figure}
  \centering
  \includegraphics[width=0.5\columnwidth]{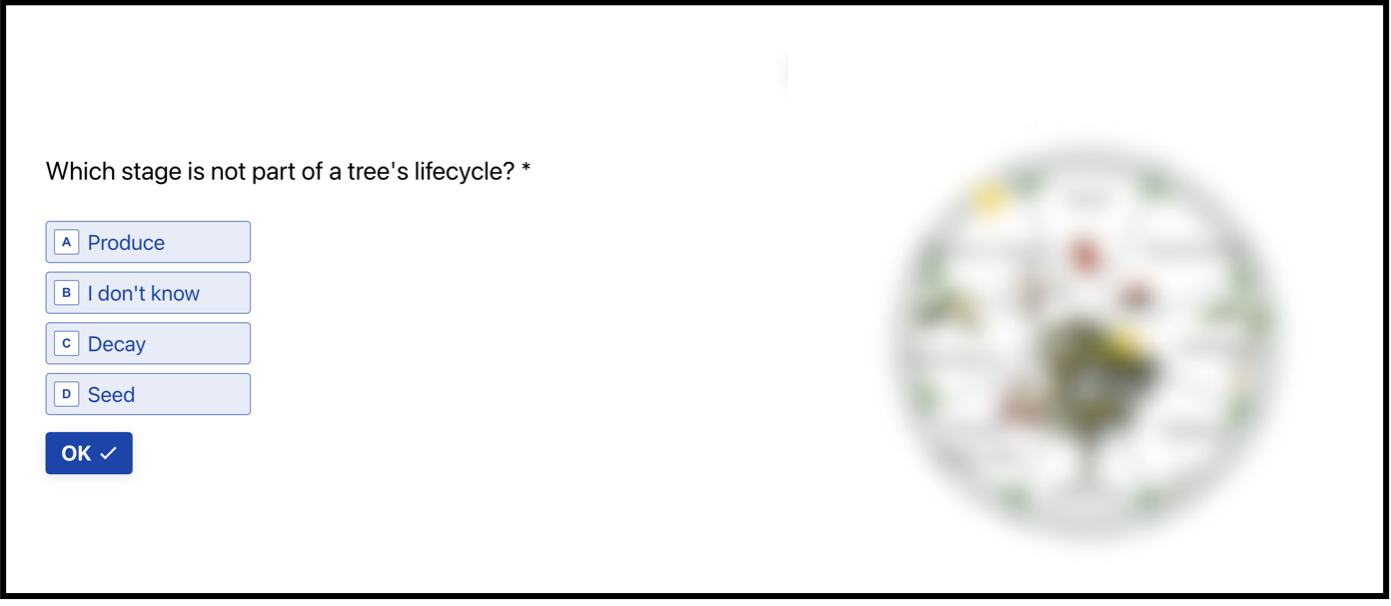}
  \caption{Information recall assessments interface.}
  \label{fig:recall}
\end{figure}

\textbf{Pre-Study Baseline Assessment.}
The first session aims to obtain the participants' initial level of knowledge prior to viewing the infographic materials. This session serves as a reference point for measuring the acquisition and retention of information throughout the subsequent experiment sessions. 

During this phase, the participants completed a fact-based multiple choice questionnaire related to the concepts of the selected infographics. An example of the questionnaire is included in \autoref{fig:baseline} for reference. There are a total of 15 questions, corresponding to the 15 selected infographics. Each question had four choices, including one correct option, two incorrect options, and an ``I don't know'' option. Upon completion of the questionnaire, we measured the correctness in the responses, which was subsequently converted into an accuracy score.

\textbf{Comprehension Assessment.}
In the second session, the comprehension assessment phase, we aimed to assess the participants' level of understanding after exposure to infographic stimuli. Each infographic stimulus was individually displayed on the screen. While viewing the infographic, participants were asked to write a brief summary of its main idea or theme. This step was implemented to ensure that participants viewed and mentally processed the information in the infographic, rather than merely skimming and skipping it casually. Subsequently on the next screen, participants were faced with a fact-based multiple-choice question tied to the information from the preceding infographic. It's important to note that each infographic stimulus was only shown once, and participants were not allowed to revisit it.

Following the question, participants were asked to rate their confidence in their answer using a 9-point Likert scale. Additionally, they were asked to assess their emotional response to the infographic using a 9-point pleasantness scale drawn from the Self-Assessment Manikin (SAM) valence scale \cite{bynion2020self}. This comprehensive approach allowed us to evaluate not only the accuracy of participants' responses but also their perceived understanding of the information and their emotional response to the different infographics. The interfaces of the comprehension assessment session is illustrated in \autoref{fig:comprehend}.

\textbf{Short-term Information Recall.}
The short-term information recall was evaluated three days post the presentation of the infographic stimuli, serving as a follow-up to the initial assessment. In this session, participants answered a reworded version of the same set of questions asked previously. Each question screen showed a blurred version of the corresponding infographic, providing no visual details or textual content. The use of blurred images served as a brief reminder of the infographic and the topic of the questions without providing specific details. This approach facilitated the assessment of information retention over a short period. The resulting scores, reflecting the participants' retained knowledge, could then be compared to their initial answers, providing an insight into memory decay over time.

\textbf{Long-term Information Recall.}
The long-term information recall was evaluated seven days after the initial presentation of the infographic stimuli. In this session, participants answered a rephrased version of the original set of questions. This test mirrored the short-term recall assessment but focused on gauging the retention of information over a longer period. The interfaces of the short-term and long-term information recall sessions are illustrated in \autoref{fig:recall}.

\section{Results}
In our infographic experiments, we gathered two distinct categories of data: objective data derived from fact-based multiple-choice responses and subjective data originating from self-assessed scales of pleasantness and confidence. To process these findings, we first numerically coded the objective data. Correct answers received a code of 1, while incorrect responses and those marked as ``I don't know'' were assigned a code of 0. Following this initial coding, we performed data transformation by aggregating the responses and organizing them according to the specific experimental conditions and individual participants. Accuracy scores are calculated as the number of correct responses divided by the total number of responses. We then proceeded to evaluate the impact of infographic exposure on comprehension and recall over time.

\subsection{Validating the Effectiveness of the Infographic Stimuli}
Before delving into condition-specific findings, we first validated the effects of infographic presentations. To accomplish this, we analyzed changes in accuracy scores at four distinct time points: baseline (before infographic presentation), comprehension (immediately after), short-term recall (three days after), and long-term recall (seven days after).

\begin{figure}
  \centering
  	\includegraphics[width=0.7\columnwidth]{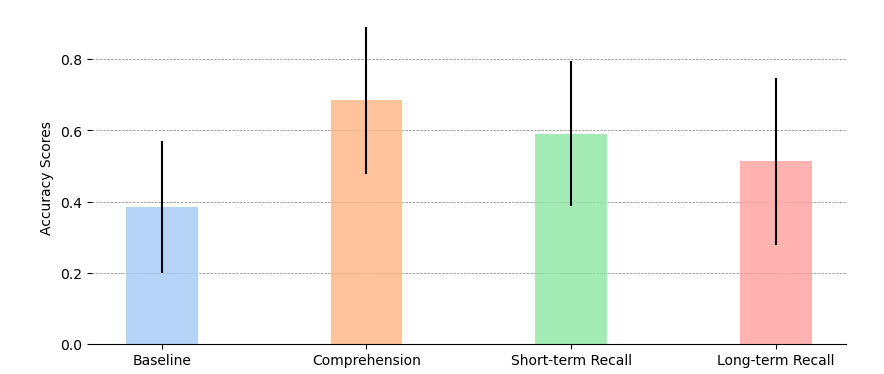}
  	\caption{Results of the average accuracy score of the pre-study baseline assessment before exposure to the infographic stimuli, the comprehension assessment immediately after the infographic presentation, the short-term recall assessment three days after the infographic stimuli onset, and the long-term recall assessment seven days after the infographic stimuli onset.}
  	\label{fig:time}
\end{figure}

The changes in the accuracy scores are illustrated in \autoref{fig:time}. The pre-study baseline score was 0.384 (SD=$0.185$), representing participants' intrinsic knowledge before being exposed to the infographics. Following the infographic presentation, the mean accuracy score increased to 0.684 (SD=$0.208$), suggesting that the visual stimuli were effective in conveying information and that participants successfully grasped the main ideas. Subsequent assessments of short-term and long-term recall aimed to evaluate information retention over time. In line with Ebbinghaus's psychological theory of the human forgetting curve \cite{ebbinghaus1885gedachtnis}, accuracy scores diminished with time, resulting in short-term recall accuracy of 0.591 (SD=$0.205$) and long-term recall accuracy of 0.513 (SD=$0.236$).

We employed a repeated-measures ANOVA to compare the statistical differences between these time-separated measurements. The analysis revealed a significant difference between the four groups ($p\textless0.01$ with Greenhouse-Geisser correction). Furthermore, post-hoc pairwise comparisons using Bonferroni correction indicated significant differences among all four groups ($p\textless0.01$ for all comparisons). 
Overall, these results affirm the effectiveness of our infographic design in conveying information and validate the user assessment procedure.

\subsection{Investigating Perceptual Effects Using Mean Scores and Change Scores}

To examine the intricacies of text placement and descriptiveness and their effects on various perceptual metrics, we utilized and visualized two distinct score types: mean scores and change scores.

\textbf{Mean scores}, defined as the arithmetic average of scores within a given metric, serve to indicate the central location of sample data. In the context of our study, these scores provide a measure of participants' knowledge level or perceptual experiences. We present data regarding participants' pleasantness, confidence, and comprehension in the form of these mean scores. Pleasantness scores are standalone metrics representing the degree of enjoyment participants experienced while viewing the infographics. Similarly, Confidence scores provide a separate measure, reflecting the level of confidence participants felt in the accuracy of their responses. The Comprehension score, meanwhile, reflects the knowledge participants retained immediately after the infographic presentation.

In contrast, \textbf{change scores} are determined by the difference in an individual's scores across two or more distinct time points, providing a valuable tool for assessing the impact of experimental interventions \cite{rogosa1982growth}. In our case, we utilized change scores to assess two key aspects: information gain and memory loss. Information gain was measured by calculating the differences between the comprehension score and the pre-study baseline score. This metric quantified the extent of knowledge gained from the initial infographic presentation, providing an indicator of the magnitude of knowledge gained. In tandem with this, we evaluated memory loss by examining both short-term and long-term recall. Short-term memory loss, defined as the difference between the short-term recall accuracy (recorded on day 3) and the comprehension score, aimed to quantify the degree of information forgotten over a relatively brief period. Similarly, long-term memory loss was determined by the difference between the long-term recall accuracy (recorded on day 7) and the comprehension score, providing an indication of the extent of information loss over a longer time frame.

\begin{table}
  \caption{Outcomes of the split-plot ANOVA, bolding indicate significance.}
  \label{tab:ANOVA}
  \scriptsize%
  \centering%
  \begin{tabular}{m{2.2cm}cccccc}
  \toprule
 & \adjustbox{angle=60, valign=t}{Pleasantness} & \adjustbox{angle=60, valign=t}{Confidence} & \adjustbox{angle=60, valign=t}{Comprehension} & \adjustbox{angle=60, valign=t}{\shortstack{Information \\ Gain}} & \adjustbox{angle=60, valign=t}{\shortstack{Short-Term \\ Memory Loss}} & \adjustbox{angle=60, valign=t}{\shortstack{Long-Term \\ Memory Loss}} \\
  \midrule
\shortstack[l]{Placement} & 0.455 & 0.259 & 1.00 & 0.485 & \textbf{0.045} & \textbf{0.018} \\
\shortstack[l]{Descriptiveness} & \textbf{0.018} & 0.397 & 0.703 & 0.440 & 0.512 & 0.435 \\
\bottomrule
\end{tabular}
\end{table}

\begin{figure}[htbp]
  \centering
  	\includegraphics[width=0.5\columnwidth]{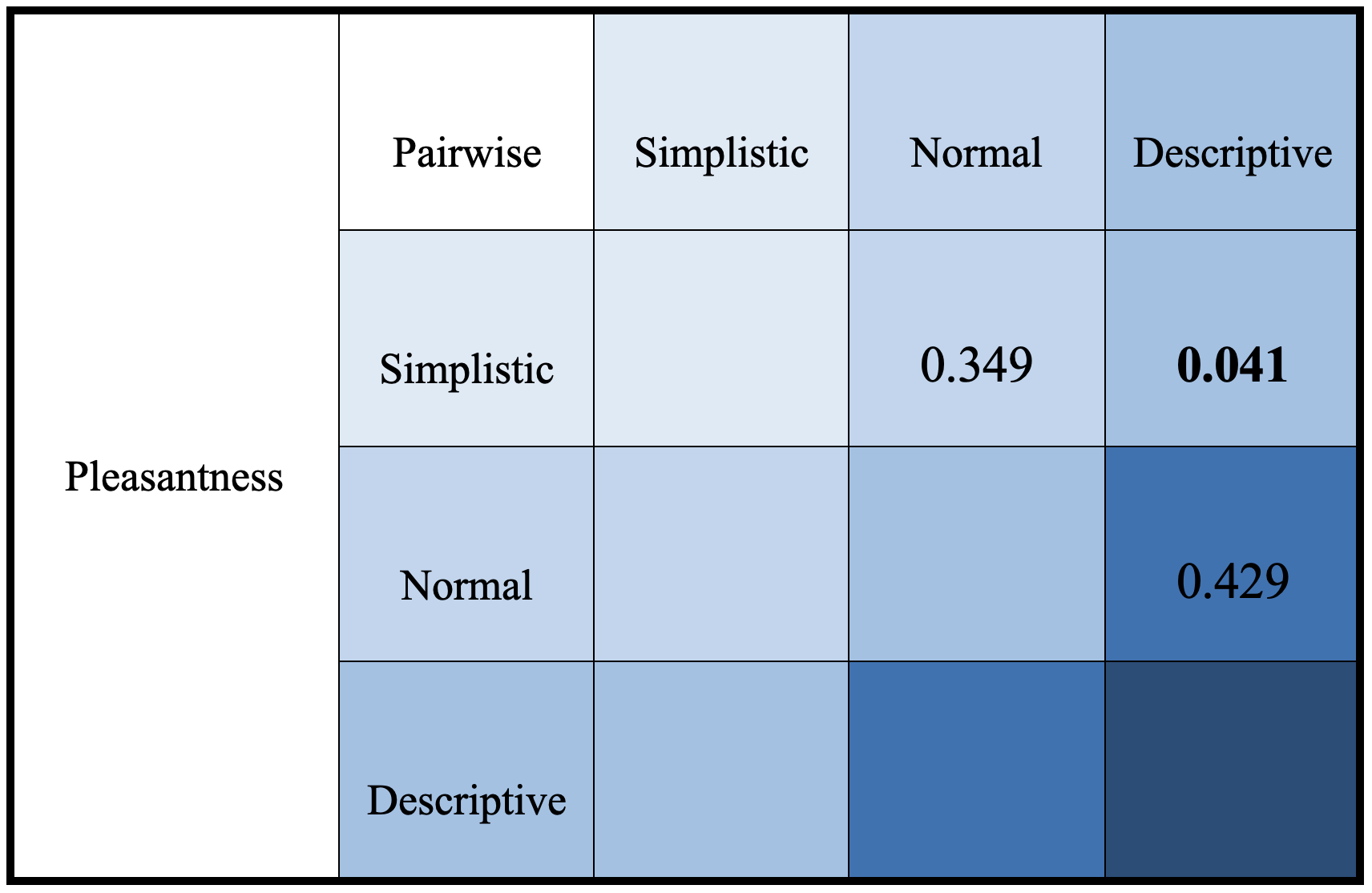}
  	\caption{Pairwise comparison results of the descriptiveness levels on pleasantness}
  	\label{fig:pairwise}
\end{figure}

\begin{figure*}[htbp]
  \centering
  \includegraphics[width=\textwidth]{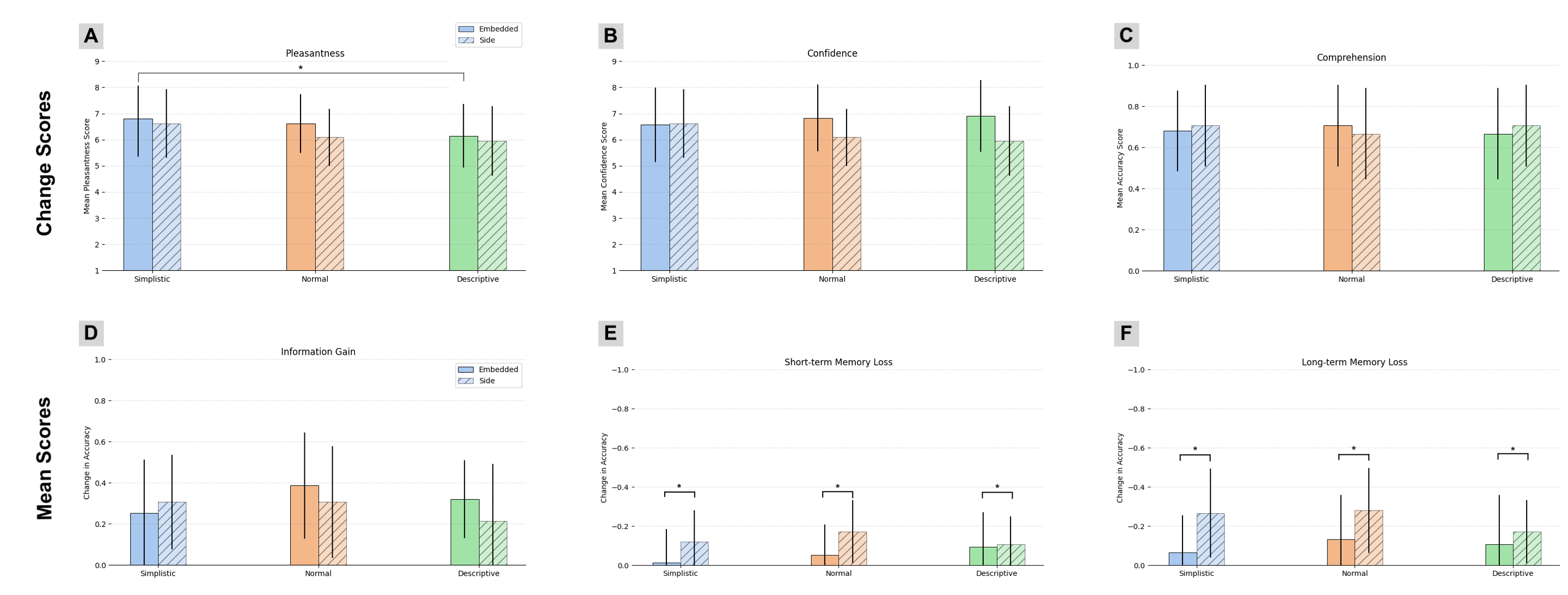}
  	\caption{Results of the mean scores and change scores represented in a grouped bar chart. The unshaded bars represent the embedded text condition whereas the shaded bars represent the side text condition. The bars are grouped according to the three levels of text descriptiveness, namely simplistic, normal and descriptive. A: the pleasantness score, B: the confidence score, C: the comprehension score, D: the information gain after viewing the inforgraphic stimuli, E: the short-term memory loss, calculated as the difference between short-term recall accuracy and comprehension score, F: the long-term memory loss, calculated as the difference between long-term recall accuracy and comprehension score. We identified statistical significance in A, E and F. * indicates $p<0.05$. }
  	\label{fig:combined}
   \vspace{-4mm}
\end{figure*}

To evaluate the statistical impact of each experimental condition, we utilized a split-plot mixed ANOVA. Text placement served as the between-subject factor, while text descriptiveness functioned as the within-subject factor. \autoref{tab:ANOVA} encapsulates the outcomes of this analysis. Our results identified a significant effect between text placement and both short-term ($p=0.045$) and long-term memory loss ($p=0.018$). In terms of the level of descriptiveness, a significant relationship was found between text descriptiveness and pleasantness ($p=0.018$). However, no interaction effects were detected. We did not identify any statistically significant difference in confidence, comprehension or information gain between the experimental conditions.

Given that text descriptiveness encompasses three levels, we conducted a post-hoc test with Bonferroni correction specifically for the pleasantness aspect. This resulted in a pairwise comparison chart, depicted in \autoref{fig:pairwise}. We identified a significant difference in pleasantness between the simplistic and descriptive conditions ($p=0.041$).

Collectively, the results derived from the six experimental conditions ($2$ \textbf{Text Placement} $\times$ $3$ \textbf{Text Descriptiveness}) are depicted in \autoref{fig:combined}, summarizing the three mean scores and three change scores.

\section{Discussion}
\subsection{Simplicity and Pleasantness}

Our analysis revealed that text descriptiveness significantly influenced the degree of pleasantness experienced by participants when viewing infographic stimuli. Interestingly, as displayed in \autoref{fig:combined}A, we observed that the level of pleasantness tended to diminish as the text became more descriptive. This trend suggests a linear relationship, as supported by a test of within-subjects contrasts ($F=6.939,p=0.014$) within the framework of the split-plot ANOVA. These findings imply that infographics employing simpler texts may enhance the viewing pleasantness, while those featuring more descriptive and extensive texts could potentially detract from the overall experience.

Our findings align well with general design principles that advocate ``less is more'', as well as theoretical concepts of simplicity in information visualization, such as Tufte's data-ink ratio \cite{tufte1990data, inbar2007minimalism}. This principle emphasizes maximizing the use of ``ink'' to represent data and essential information while minimizing the use of ``ink'' for textual embellishments and other non-essential elements. In the context of our study, simpler texts contribute to minimizing the overall use of ``ink'', thereby improving the data-ink ratio. This, in turn, potentially enhances the pleasantness and enjoyment derived from viewing infographics. While the concept of data-ink ratio can be a subject of debate, it serves as a basic guideline for evaluating simplicity in design.

The cognitive load theory \cite{sweller1991evidence} also offers a potential explanation for the influence of text descriptiveness on perceived pleasantness. As the descriptiveness and length of the text increase, the cognitive load required to process the information contained in an infographic could also increase. This increasing demand on cognitive resources could subsequently diminish the viewer's sense of enjoyment or pleasantness. Therefore, this theory aligns with our observation that simpler text can lead to a more pleasant experience, as it potentially reduces the cognitive load, making the infographic easier to process.

The term ``simplicity'' encompasses a variety of factors. In our study, we manipulated the level of descriptiveness primarily by modifying the length of textual statements, while keeping the fundamental message intact. As such, we employ the term ``simplistic'' to describe text that is shorter in length but still carries essential information. However, it is crucial to acknowledge that simplicity or complexity cannot be solely determined by length. Various semantic and lexical factors, such as specific wording, vocabulary rarity, writing style, sentence structure, and so more, can contribute to the overall perception. Consequently, our interpretation of the impact of simplicity on pleasantness is a relatively narrow focus. The multifaceted concept of simplicity presents ample opportunities for further exploration in future research endeavors.

Lastly, it should be noted that a significant difference in pleasantness was only observed between the simplistic and the descriptive conditions ($p=0.041$). This observation suggests that as a perceptual metric shaped by human judgment, pleasantness exhibits a certain degree of tolerance with respect to levels of text descriptiveness, only responding to more substantial variations in these conditions.

\subsection{Text Embedding and Memory Retention}

Upon examining the impact of text placement on memory retention, we observed a significant difference between embedded text and side text, as demonstrated in \autoref{fig:combined}E and F. Specifically, conditions involving embedded text resulted in significantly less short-term ($p=0.045$) and long-term memory loss ($p=0.018$). These findings suggest that the embedding of text within infographics potentially enhances memory retention when compared to placing text alongside the visuals.

These results are consistent with theoretical expectations and can be explained by the split-attention effect as part of the cognitive load theory. The role of split-attention effect has been extensively investigated in learning \cite{sweller1991evidence, mayer1998split} and situated analytics \cite{Wen:2022:EVL}. In our case, the side text condition may have split the participant's attention to focus on the text and visuals separately, requiring additional mental power for integrating the information together once again. On the other hand, the embedded text condition creates a unified source of information by integrating text elements within the visual components, thus mitigate the split-attention effect and promotes memory retention and learning.

\subsection{Consistent Comprehension In Varying Text Conditions}
Interestingly, our study did not find any statistically significant differences in comprehension between the varying levels of text conditions. The comprehension and information gain scores remained within a consistent range with only small statistical variations (\autoref{fig:combined}C, D).

This observation could be attributed to several factors. First, the experimental conditions we selected might not have been sufficiently diverse to induce perceptible differences in comprehension. Secondly, text placement and descriptiveness might not be the principal factors dictating immediate comprehension and information transfer.

However, it's crucial to note that this finding does not diminish the significance of text-visual interplay in effective communication. In fact, it opens up avenues for future research to investigate the influence of other text-visual aspects, such as the role of font size and typographical emphasis on comprehension.

\subsection{Infographic Design Implications}
Our observations on text descriptiveness and text placement's impact on pleasantness and memory retention offer several important implications for infographic design.

\subsubsection{Simplicity in text can improve pleasantness experience.} As indicated by our experimental results, simplistic text design contributes to statistically higher pleasantness scores, indicating an overall improved user experience. When designing for pleasantness, designers and researchers should consider incorporating more simplistic textual elements and avoiding excessively lengthy and verbose explanations. While our study primarily focused on the length of individual text statements as a measure of descriptiveness, it's worth noting that simplicity extends far beyond this dimension, encompassing a wide array of semantic and lexical factors. 

In designing infographics, striking a balance between providing sufficient context and minimizing the use of text is crucial. The criteria of ``simplistic'' text and whether enough context is provided can vary, depending on the infographic's purpose and its target audience. For example, more complex or domain-specific infographics may require more extensive text to provide necessary context information. Hence, designers should consider the context and audience for an infographic and adjust their simplicity criteria accordingly.

It's also important to note that infographics can encompass different types of textual information, including but not limited to titles, captions, and data explanations. Each of these text types may underscore different considerations in terms of descriptiveness. A title, for example, may need to be succinct and direct, while background descriptions might require more elaborate explanations.

In summary, designers and researchers can employ simpler texts in infographics design for an optimal pleasantness experience. This strategy, however, should be employed with a clear understanding of the specific purpose and audience of the infographic, and with an awareness of the different requirements of various text components.

\subsubsection{Embedding text within visual elements can improve memory retention.}
Our results illustrate that text placement can have a significant effect on both short-term and long-term memory retention. In particular, embedding text within visual elements appears to reduce information loss over time compared to placing text adjacent to these components. This advantageous outcome may be attributed to the mitigation of the split-attention effect and the reduction of cognitive load. As such, when designing for more memorable infographics that promote long-lasting impressions, designers may consider text embedding as an effective layout strategy.

While embedded text has shown promising results in our study, certain practical considerations call for attention when implementing this approach. The effectiveness of embedded text placement is contingent on the design of the graphic elements and the density of the text. Optimal use of the limited space offered by a circular visualization requires a delicate balance between graphics and text. Overcrowding a small area with dense embedded text may compromise the clarity and readability of the information, thereby counteracting the benefits of this placement strategy. Furthermore, intricate graphical elements risk being covered and obscured by the embedded text, imparing the visual processing of the infographic. Thus, when considering the use of embedded text in practice, additional factors that may moderate its effectiveness should be taken into account.

Furthermore, the design of the textual elements themselves can have an impact on the memorability of the infographics. While our study did not investigate this aspect, it's reasonable to hypothesize that visual attributes of the text, such as color, density, alignment, and font size, can impact the quality of the text embedding and, by extension, the overall effectiveness of the infographic. Therefore, these design features demand careful consideration and further exploration when embedding text in practice.

Lastly, while our study found promising results for text embedding, it's important to acknowledge the limited generalizability of these findings to other types of visualizations. Our focus was predominantly on circular infographics, such as pie charts and donut charts, which provide ample open space for text embedding within the visual elements. However, the applicability of text embedding may not extend as readily to other visualization types. For instance, scatter plots or line charts may lack sufficient space within their visual elements to accommodate text embedding. Even for visualizations where area is the main visual encoding, such as mosaic charts and bubble charts, the effectiveness of text embedding in enhancing memorability remains an open question. Consequently, we urge designers and researchers to consider these limitations and pursue further exploration. It's crucial to understand that the success of text embedding in different visualization contexts may vary and require different design approaches. Ultimately, the goal is to tailor the design strategy to the specific requirements and constraints of each visualization type.

In summary, the findings of our study provide some guiding principles for circular infographic design. 

\vspace{1em}

However, the design space of text-visual interplay in infographics design has vast potential for further exploration. For instance, the strategies for text embedding and descriptiveness can vary significantly across languages. Alphabetic languages and pictographic languages exhibit distinct visual layouts and spatial coverage. Chinese text, for instance, is often more compact compared to English when conveying the same sentence. This is because a single Chinese character represents a complete morpheme, while a letter in English only denotes part of a morpheme. Therefore, the criteria for descriptiveness might need adjustment when designing for different languages. Furthermore, text position and orientation could also affect user perception. Building upon the split-attention effect, we hypothesize that text labels placed distant from their associated visual elements might lead to suboptimal memorability performance. We encourage designers and researchers alike to delve into the open-ended questions surrounding these methods and extend the investigation to other types of visualizations. It is through this collective curiosity and innovation that we can enhance our understanding and continuously refine the art and science of infographic design.

\section{Conclusion}
In this study, we investigated the effects of text placement and descriptiveness on the pleasantness, comprehension, and memorability of circular infographics. We created six unique experimental conditions by systematically manipulating the placement (embed, side) and descriptiveness (simplistic, normal, descriptive) of text within fifteen infographic materials. We recruited thirty participants and evaluated their pleasantness, comprehension and memory retention using both mean scores and change scores on accuracy scores and self-reported perceptual experience. 

Our empirical results suggested statistically significant findings that simpler and less descriptive text generally led to a more pleasant viewing experience. Moreover, we found that the strategy of embedding text within the infographic positively influenced memory retention when compared with placing the text alongside the visual elements. These findings underscore the importance of design decisions in text layout in infographic design. We provided guiding design implications for circular infographic design that advocate for simpler, embedded text for enhanced pleasantness and memorability. However, the design of infographic is a complex procedure that involves various perceptual intricacies. Careful considerations of the purpose and audience of the target infographic, as well as more detailed text design attributes are needed when employing these strategies.

Despite these insights, our study has some limitations. First, the robustness of the results could be improved by including more levels of experimental conditions. For example, more variations of the descriptiveness levels could be included to encapsulate a wider range of text complexities. Second, additional user experience metrics, such as engagement and accessibility, can be explored for a holistic understanding of how different design decisions impact the perceptual effectiveness of infographics. Third, implicit measures, such as response time, can be included as evaluation measures to provide additional insights into the learnability of infographics.

Our present study involves manual edits of infographic materials into different text layouts. However, future work could benefit from adopting automatic infographic generation strategies\cite{cui2019text,cui2021mixed}. This would enable the creation of more controlled and consistent infographic stimuli in large quantities, suitable for robust, large-scale studies. Moreover, this method could be beneficial when testing generalizability across different types of visualizations. By developing an informative database encompassing a variety of topics and distinct visualization charts, we can systematically evaluate the perceptual effects across different domains and visualization types.

As the field of infographic design continues to evolve, it is our hope that our findings will inform and inspire further research and innovation in this area.

\section*{Acknowledgement}
This work was supported by the National Natural Science Foundation of China 62272396.

\bibliography{sn-bibliography}

\end{document}